\documentclass[showpacs,twocolumn,aps,floatfix,superscriptaddress]{revtex4-1}
\usepackage{amsmath,amssymb,eucal,graphicx,subfigure,color}
\begin{document}



\def\b{\beta}
\def\blead{\beta_{\rm lead}}
\def\blag{\beta_{\rm lag}}
\def\r{{\bf r}}
\def\Plead#1{P_{\rm lead}(#1)}
\def\Plag#1{P_{\rm lag}(#1)}
\def\av#1{\langle#1\rangle}

\title{Ordering statistics of 4 random walkers on the line}

\author{Brian Helenbrook}
\email{helenbrk@clarkson.edu}
\affiliation{Mechanical \& Aeronautical Engineering, Clarkson University, Potsdam, NY 13699-5725}
\affiliation{Department of Mathematics \& Computer Science, Clarkson University, Potsdam, NY 13699-5815}
\affiliation{Clarkson Center for Complex Systems Science (C$^{\,3}$S$^{\,2}$), Clarkson University, Potsdam, NY 13699}
\author{Daniel ben-Avraham}
\email{qd00@clarkson.edu}
\affiliation{Physics Department, Clarkson University, Potsdam, New York
13699-5820} 
\affiliation{Department of Mathematics \& Computer Science, Clarkson University, Potsdam, NY 13699-5815}
\affiliation{Clarkson Center for Complex Systems Science (C$^{\,3}$S$^{\,2}$), Clarkson University, Potsdam, NY 13699}

\begin{abstract}
We study the ordering statistics of 4 random walkers on the line, obtaining a much improved estimate for the long-time decay exponent of the probability that a particle leads to time $t$; $P_{\rm lead}(t)\sim t^{-0.91287850}$, and that a particle lags to time $t$ (never assumes the lead); $P_{\rm lag}(t)\sim t^{-0.30763604}$.  Exponents of several other ordering statistics for $N=4$ walkers are obtained to 8 digits accuracy as well.  The subtle correlations between $n$ walkers that lag {\em jointly}, out of a field of $N$, are discussed: For $N=3$ there are no correlations and $P_{\rm lead}(t)\sim P_{\rm lag}(t)^2$.  In contrast, our results rule out the possibility that $P_{\rm lead}(t)\sim P_{\rm lag}(t)^3$ for $N=4$, though the correlations in this borderline case are tiny. 
\end{abstract}

\maketitle

\section{Introduction}
Imagine $N$ random walkers on the line, each stepping to the right or left at equal rates (or all diffusing with the same diffusion constant $D$), initially started at locations $x_1(0)<x_2(0)<\cdots<x_N(0)$.  For the case of ``vicious walkers," the  process terminates as soon as any two walkers cross one another, thus violating the initial ordering.  The probability that the vicious walkers process lasts to time $t$ decays asymptotically as $t^{-\beta_N}$,
with $\beta_N=N(N-1)/4$~\cite{fisher84,huse84}.  The algebraic decay with time is typical of the survival of other kinds of ordering.  For example, for the ``leader" problem, the probability that the leading particle remains in the lead at all times, i.e., that  $x_1(t)<x_i(t)$, $i=2,3,\dots,N$ (regardless of the ordering of the remaining particles), decays also as $t^{-\beta_N}$, but with different values of the exponent $\beta_N$~\footnote{There is no accepted consensus on the notation for the exponents pertaining to the various orderings.  We use the notation $\beta$ for all cases, making the distinction clear by context.}.
In this case, only $\beta_2=1/2$ and $\beta_3=3/4$~\cite{niederhausen83,fisher88,dba88,redner01}, as well as the limit $\beta_N\sim(\ln N)/4$
as $N\to\infty$~\cite{kesten92,krapivsky96,redner99,dba03} are known exactly.  For the ``laggard" problem, the asymptotic probability that particle $i$ ($i>1$) never assumes the lead ($x_i\nless x_1,\dots,x_{i-1},x_{i+1},\dots,x_N$), is known exactly only for $\beta_2=1/2$, $\beta_3=3/4$ and for $\beta_N\sim(\ln N)/N$ in the limit of $N\to\infty$~\cite{dba03}, etc.

For $N=2$ the probability that the two particles retain their original ordering up to time $t$ decays as $t^{-1/2}$.
For $N>2$, the coordinates $x_1(t),x_2(t),\dots,x_N(t)$ may be regarded as representing a {\em single} random walker in $N$-dimensional space~\cite{fisher88}.  In this representation, the constraint that particles $i$ and $j$ never cross corresponds to the surface $x_i=x_j$: as long as the walker remains to one side of that surface the ordering between the two particles is conserved.  Motion of the single walker along the axis $x_1=x_2=\cdots=x_N$ does not affect the distances between the original particles, thus for ordering statistics it suffices to focus on the $(N-1)$-dimensional subspace perpendicular to that axis. The 2-dimensional subspace for the case of $N=3$ is shown in Fig.~\ref{n3.fig}.  Each of the six wedges in the figure represents a particular ordering of the particles ---
for example, the wedge labeled `132' corresponds to the ordering $x_1<x_3<x_2$ --- and crossing any of the walls
(or lines, in the perpendicular subspace) results  in a reversal of the ordering of the corresponding particles.  Vicious walks correspond to the case that the single walker remains confined to the `123' wedge. For the leader problem,
the single walker must remain within the {\em adjacent} `123' and `132' wedges, etc.  It is clear from these considerations that for $N=3$ there are only 6 types of ordering statistics, corresponding to the number of adjacent wedges that the single walker is allowed to visit~\footnote{One could think of more involved statistics, though.  For example, all wedges are allowed, but crossing {\it directly} between `123' and `132' is disallowed ($\beta=1/4$); or crossings might be allowed only in a particular sequence of events, etc.  Here we limit discussion to the simplest case that if adjacent wedges are allowed, so are the crossings between them.}:

\begin{figure}[t]
\includegraphics[width=0.35\textwidth]{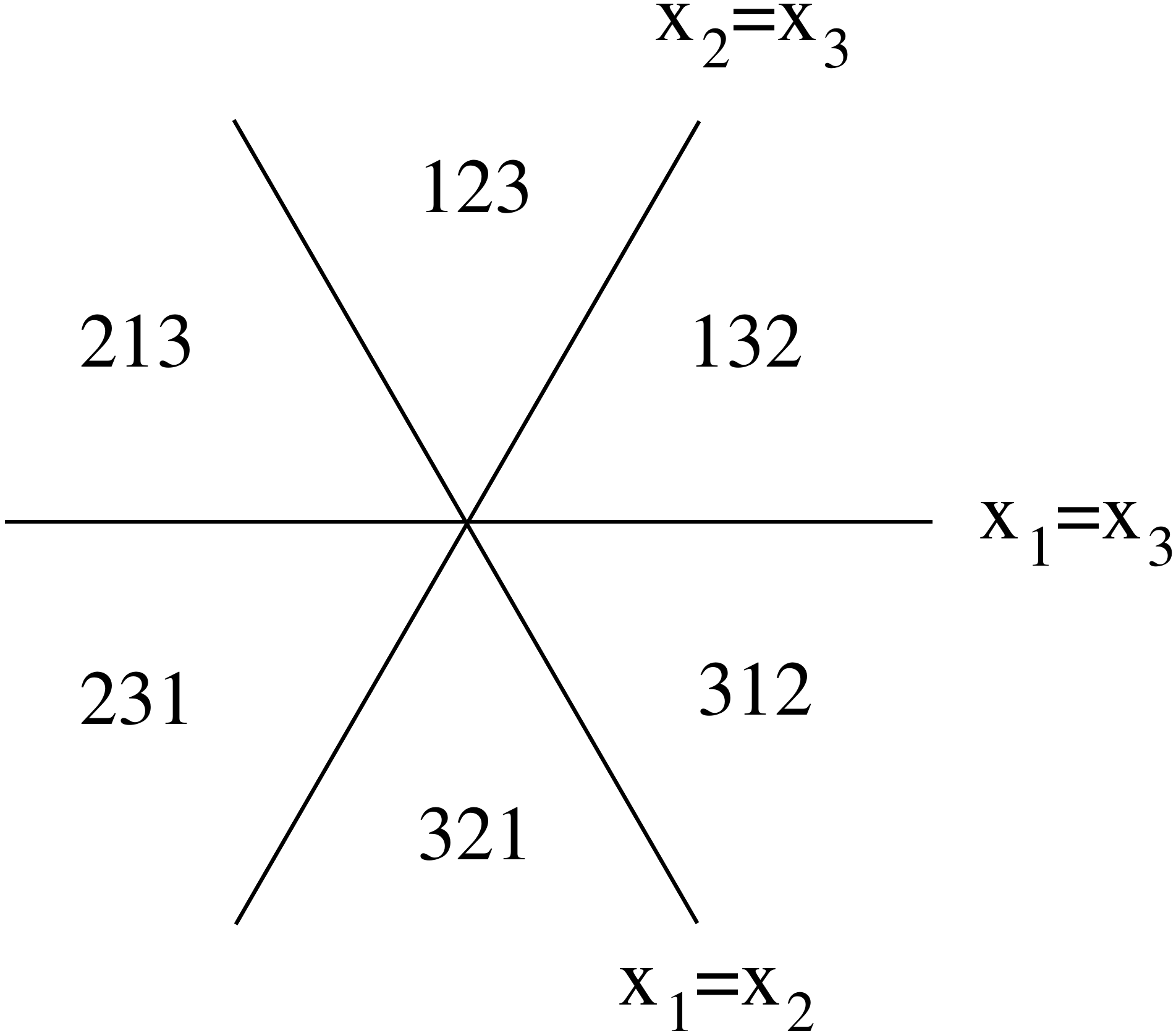}
\caption{Two-dimensional subspace for the single walker representing the case of $N=3$. The $x_i=x_j$ surfaces appear as lines in this subspace.  The six wedges demarcated by these lines are labeled according to the ordering of the 3 particles in the original problem.
}
\label{n3.fig}
\end{figure}

\begin{itemize}

\item[]{1 wedge:} This is the case of vicious walkers, for which $\beta=3/2$.

\item[]{2 wedges:} This is the ``leader" problem, that 1 remains ahead (to the left) of 2 and 3.  The decay exponent is $\beta=3/4$.

\item[]{3 wedges:} The particles may be in any of the orderings `123', `213', or '231', say.  Put differently, 2 must remain ahead of 3, while the location of 1 is irrelevant.  This is the ordering statistics for $N=2$, with $\beta=1/2$.

\item[]{4 wedges:} The ``laggard" problem, where a particle is never allowed to assume the lead, for which $\beta=3/8$.

\item[]{5 wedges:} The particles are not allowed to meander into a particular ordering (`321', say) but are allowed all other orderings.  In this case $\beta=3/10$.

\item[]{6 wedges:} The trivial case, where the particles are allowed to freely explore all orderings.  In this case the process never terminates, so $\beta=0$.

\end{itemize}
In other words, the case of $N=3$ is well understood.  (This remains true even for the more general case that the walkers have different diffusion constants~\cite{dba88,dba03}.)  In contrast, relatively little is known for $N>3$. 

\begin{figure}[b]
\hskip0.08\textwidth \includegraphics[width=0.35\textwidth]{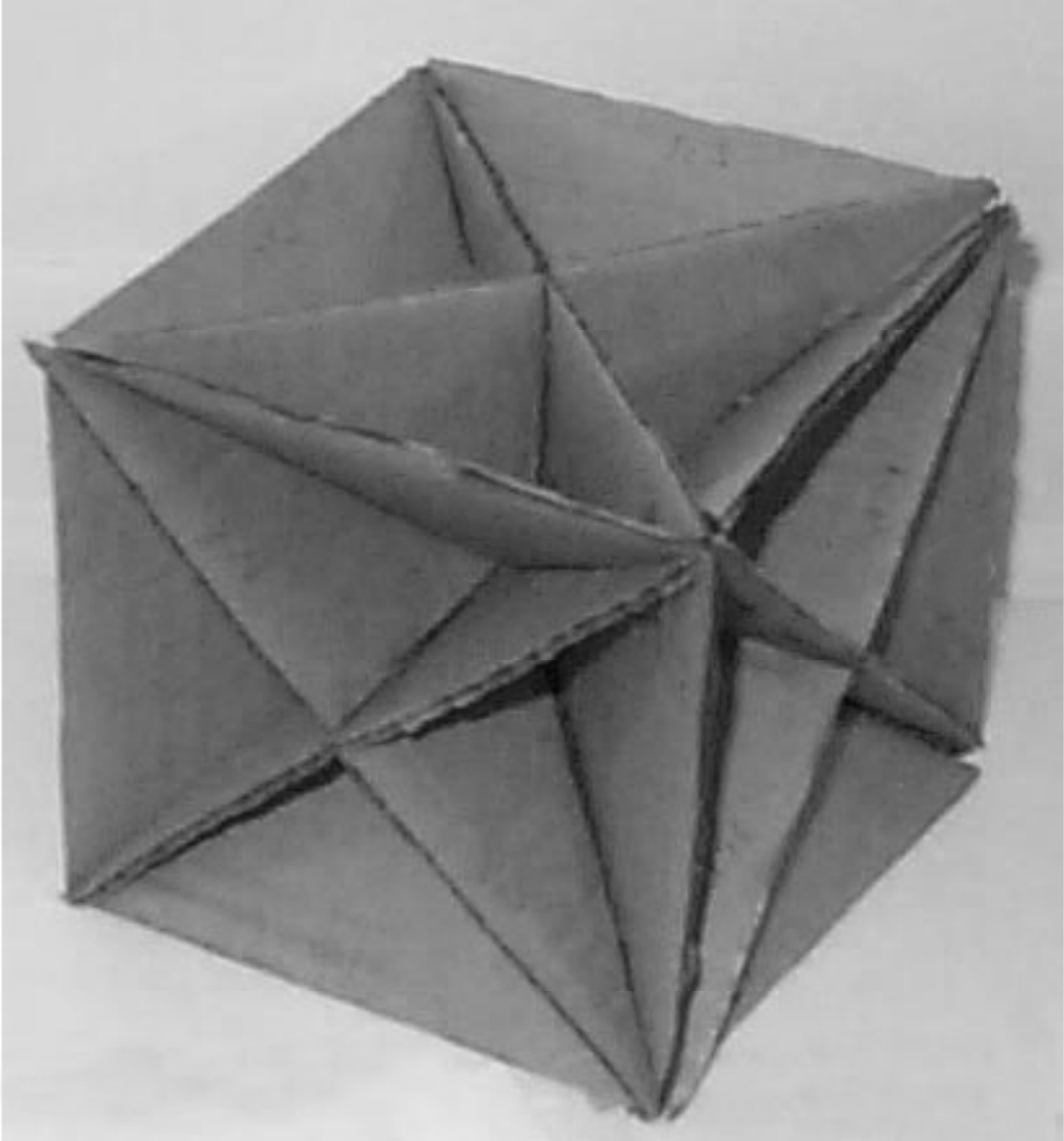}
\includegraphics[width=0.5\textwidth]{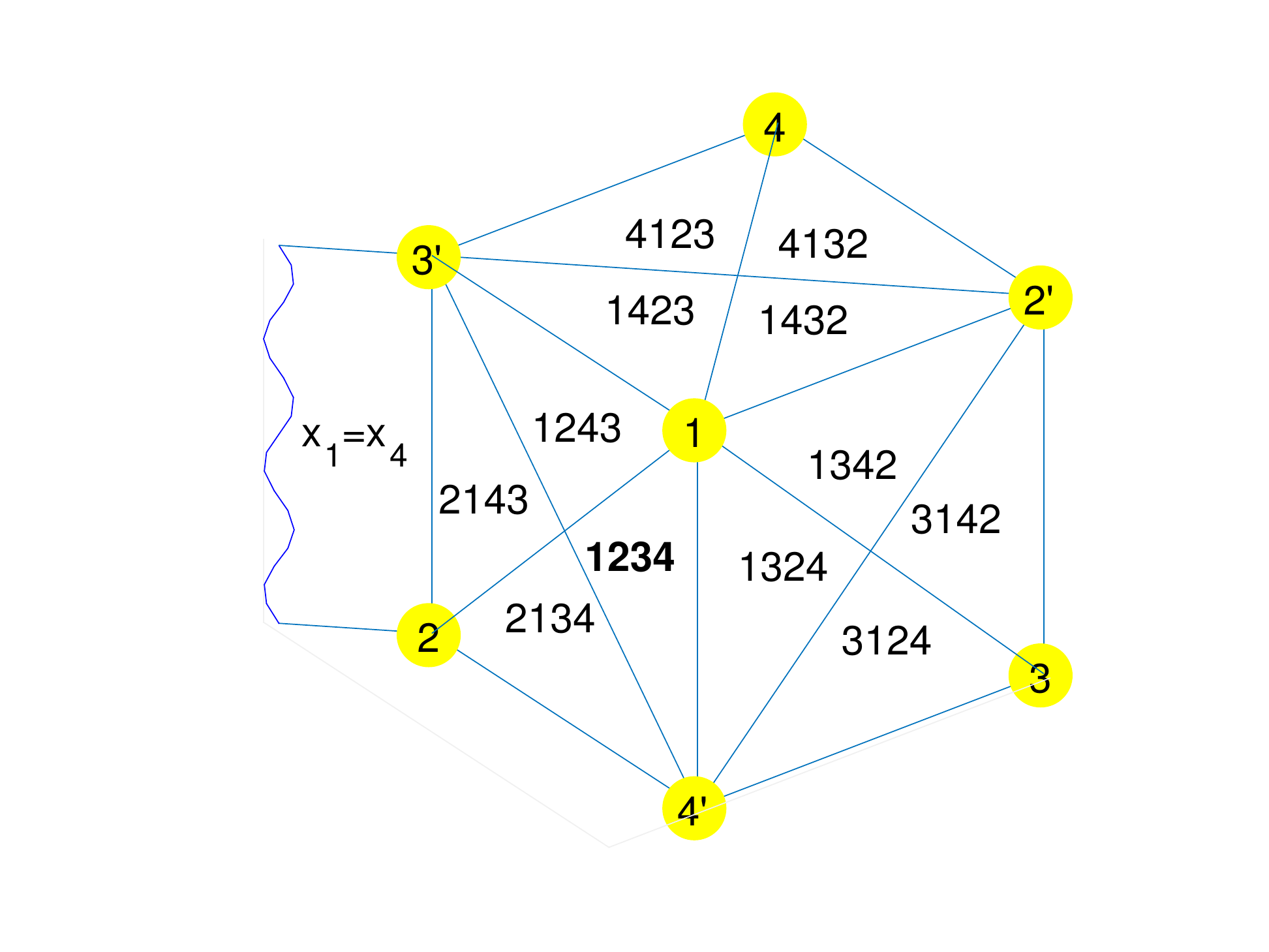}
\caption{Three-dimensional subspace of the single walker for $N=4$. Top: A cardboard model showing how the six constraint walls $x_i=x_j$ divide the space into 24 wedges.  The walls extend to infinity but are shown only within the confines of a cube, to highlight their orientations and symmetries. Bottom: Schematic redrawing of the 24 wedges.  Each wedge is labeled with its pertinent ordering and the `1234' wedge (with the original ordering) is highlighted in bold font. The wall $x_1=x_4$ is partially extended: crossing this wall particles 1 and 4 reverse their order.  The combined  six wedges around  vertex $i$ are those where particle $i$ leads; around $i'$ particle
$i$ is last ($i$ is the rightmost particle).  Only 12 of the wedges are visible in the figure.  The back branch of wedge `$ijkl$', corresponds to the reverse ordering `$lkji$'.
}
\label{n4.fig}
\end{figure}

For $N=4$ the orthogonal subspace for the single walker is three-dimensional, the various ordering statistics corresponding to semi-infinite triangular pyramidal wedges, or combinations of adjacent wedges (Fig.~\ref{n4.fig}).  Apart from the case of vicious walkers (which can be solved by the method of images) and some trivial ``degenerate" statistics (e.g., the location of one of the particles is ignored, so that effectively $N=3$) there seem to be no other known analytical solutions, but numerical estimates of $\b$ are available for a few types of ordering statistics~\cite{dba03,ben-naim10a,ben-naim10b}.  In~\cite{ben-naim10a} Ben-Naim and Krapivsky study the general question that a specific walker (out of a field of $N$ walkers) never falls bellow rank $n$.  Their numerical simulations for $N=4$ yield $\beta_4=0.913$ for the problem of the `leader' ($n=1$), and $\beta_4=0.306$ for the `laggard' ($n=3$). (They also introduce the useful ``cone approximation" for the evaluation of decay exponents, as well as a scaling analysis of ordering statistics as a function of $n/N$.)  In~\cite{dba03} the 4-walker problem is studied by mapping it into a three-dimensional electrostatic analog and solving that problem numerically. Their result for the leader, $\beta_4=0.91342(8)$, requires numerical extrapolation, casting some doubt on the accuracy of the last few digits.  

In this paper, we study the problem of $N=4$ numerically.  The 4-walkers problem is first mapped onto a three-dimensional electrostatic analog, as in~\cite{spitzer76,krapivsky96, redner01, dba03}, and we then use an ansatz for the solution~\cite{dba03} to further reduce the problem to a two-dimensional {\em finite} domain.  The latter can be solved numerically with great accuracy, yielding 8 significant digits for the decay exponents of the various ordering statistics.  Thus, for example, we find $\beta_4=0.91287850$ for the leader, and $\beta_4=0.30763604$ for the laggard problem.  In addition, we explore several other ordering statistics for the 4-walkers problem for the first time. Finally, we explore the issue of correlations between the walkers: for $N=3$ walkers there seem to be no correlations --- the probability for two walkers to lag jointly equals the product of their probabilities to lag independently (in the long time limit).  However, we show that for $N=4$ small correlations arise and we analyze their effect.

%
%

\section{Methods}
\label{methods.sec}

Instead of solving the diffusion problem
\begin{equation}
\begin{split}
&\frac{\partial}{\partial t}P(\r,t)=D\nabla^2P(\r,t)\,;\\
&P(\r,0)=\delta(\r-\r_0)\,;\qquad P(\r,t)=0 {\rm\ for\ } \r\in\partial W\,,
\end{split}
\label{P.eq}
\end{equation}
for the single walker (starting at $\r_0$ within the wedge $W$), one can look at the analogous electrostatic problem~\cite{krapivsky96,redner01,dba03,spitzer76}:
\begin{equation}
\nabla^2V(\r)=-\delta(\r-\r_0),;\qquad
V(\r)=0 {\rm\ for\ } \r\in\partial W\,.
\label{V.eq}
\end{equation}
The single walker in~(\ref{P.eq}) survives with probability $S(t)=\int P(\r,t)\,d^d\r\sim t^{-\b}$, while the potential $V$ in~(\ref{V.eq}) falls off  at large distance $r\gg r_0$  as $V(\r)\sim r^{-\mu}$.  Because
\begin{equation}
\int_0^t S(t)\,dt\sim\int^{\sqrt{Dt}}V(\r)r^{d-1}\,dr\,,
\label{S-V.eq}
\end{equation}
we have
\begin{equation}
\label{beta-mu.eq}
\beta=\frac{2-d+\mu}{2}=\frac{3-N+\mu}{2}\,,
\end{equation}
where for the last relation we put $d=N-1$, since $W$ is in the orthogonal subspace for the single walker. The equivalence~(\ref{S-V.eq}) arises from the integral of~(\ref{P.eq})  over all time, which yields essentially Eq.~(\ref{V.eq}), and from the fact that the single walker has an effective reach of length $r\sim\sqrt{Dt}$. 
Solving the problem for $V(\r)$, rather than for $P(\r,t)$, is easier because of the absence of the time variable $t$.

For $N=3$, for example, the wedges $W$ are two-dimensional (Fig.~\ref{n3.fig}) and $V(\r)$ can be expressed in polar coordinates, as $V(r,\theta)$.
Although the equation for $V$ can be solved exactly, a further simplification is achieved with the ansatz $V(r,\theta)\sim r^{-\mu}f(\theta)$, as $r\to\infty$.
Substituting this form in Eq.~(\ref{V.eq}), we get
\begin{equation}
\frac{d^2}{d\theta^2}f(\theta)=-\mu^2f(\theta)\,,
\end{equation}
(for $\r\neq\r_0$),  with the boundary conditions
\[
f(0)=0,\qquad f(\gamma)=0\,,
\]
where $\gamma$ is the opening angle of the $W$-domain in question.
The lowest eigenfunction solution, $f(\theta)=\sin(\pi\theta/\gamma)$, yields the eigenvalue (and the sought after asymptotic behavior) 
\begin{equation}
\mu=\frac{\pi}{\gamma}\,,\qquad {\rm and\ }\qquad \beta=\frac{\pi}{2\gamma}\,,
\end{equation}
the value for $\beta$ following from (\ref{beta-mu.eq}). This method was used in~\cite{dba03} to obtain the various ordering exponents for $N=3$.  In what follows, we use the very same technique for the case of $N=4$.  Writing Eq.~(\ref{V.eq}) for $V$ in spherical coordinates and applying the ansatz $V(r,\theta,\phi)=r^{-\mu}f(\theta,\phi)$, as
$r\to\infty$, we obtain
\begin{equation}
\begin{split}
\mu(\mu-1)f(\theta,\phi)&+\frac{1}{\sin\theta}\frac{\partial}{\partial\theta}\sin\theta \frac{\partial}{\partial\theta}f(\theta,\phi)\\
&+\frac{1}{\sin^2\theta}\frac{\partial^2}{\partial\phi^2}f(\theta,\phi)=0\,.\\
\end{split}
\label{ftp.eq}
\end{equation}
The domain $S$ for this equation is the $\{\theta,\phi\}$-region of the unit sphere cut out by $\partial W$ --- the walls of the domain $W$ for the original problem of $V(r,\theta,\phi)$.
The eigenvalue problem of Eq.~(\ref{ftp.eq}) is thus subject to the boundary condition  $f(\theta,\phi)=0$, for $(\theta,\phi)\in\partial S$ (the borders of the domain~$S$).  The domain corresponding to a single wedge is shown in Fig.~\ref{eigen.fig}.  When neighboring wedges are merged, the Dirichlet boundary condition can be replaced by Neumann boundary conditions along some of the borders, to exploit the symmetry of the combined, larger domain.

\subsection*{Numerical Approach}

To solve the eigenvalue problem defined by~(\ref{ftp.eq}), a continuous $hp$-finite element method was used that is based on~\cite{Helenbrook:2001}.  We used an unstructured mesh of triangles and a polynomial space of degree $p=4$ on each element.  The polynomial space was defined by
\[
{\cal T}(p) = \left\{ span\left[x^\alpha y^\beta\right] |\ 0 \le \alpha,\beta\  \&\  \alpha+\beta \le p\right\},
\]
and was represented by the modified Dubiner basis~\cite{Dubiner:1991}.  The weak form of~(\ref{ftp.eq}) was written by multiplying the equation by $\sin\theta$ then integrating over the $\theta,\phi$ domain.   
\[
\int_S \left[ 
\lambda v \sin\theta f +\frac{\partial v}{\partial\theta} \sin\theta \frac{\partial f}{\partial\theta} +\frac{\partial v}{\partial\phi} \frac{1}{\sin \theta}\frac{\partial f}{\partial\phi}  \right] d\phi \,d\theta = 0,
\]
where $\lambda = \mu(\mu-1)$ and $v$ is a test function.  Using the global space of functions defined by ${\cal T}(p)$ on each element and the interelemental continuity constraints for both $v$ and $f$, the weak form was used to generate a discrete generalized eigenvalue problem.  

The generalized eigenvalue problem was solved using SLEPC with the Rayleigh quotient conjugate gradient algorithm \cite{petsc-user-ref,slepc-manual,slepc-toms,petsc-efficient} to obtain the smallest real eigenvalue.  The convergence tolerance was set to $1 \times 10^{-10}$.   

To reduce the errors stemming from the finite resolution of the triangular mesh, a mesh adaption routine was used  \cite{Helenbrook:2017,Helenbrook_c:2002};  after each eigenvalue solution, the truncation error in the solution for the first eigenvector was employed in a solution-based mesh adaption scheme that tried to maintain the truncation error lower than a specified tolerance over the entire mesh.  After each eigenvalue solution, the mesh was then adapted to reduce the truncation error.  This was repeated four times, for increasing numbers of degrees of freedom in the mesh ($N_{DOF}$).  In addition, for each mesh size the generalized eigenvalue problem was solved three times with different initial guesses for the eigenvector, and the variations in the output eigenvalue $\lambda$ were recorded.  Results for a case with Dirichlet boundary conditions on the top and right of the domain and a symmetry boundary condition on the bottom left (which is Case~C of Table~\ref{exponents.table} below) are shown as an example, in Table~\ref{precision.table}.  As seen from this example, different guesses for the initial eigenvector affect $\lambda$ only beyond the 8th digit.  By comparing the values for the two last mesh sizes one can also conclude that  mesh size effects have converged to better than 8-9 significant digits.

\begin{table}
\begin{center}
\begin{tabular}{rc}
$N_{DOF}$ & $\lambda$ \\
1231 & 1.77913047045631e+01  \\ 
1231 & 1.77913047378482e+01  \\
1231 & 1.77913046668994e+01  \\
86485  & 1.77912995337541e+01  \\
86485 & 1.77912994085778e+01  \\
86485 & 1.77912994552660e+01 \\
210207 & 1.77912994944804e+01  \\
210207 & 1.77912994029586e+01  \\
210207 & 1.77912993511336e+01 \\
268591 & 1.77912994472011e+01 \\
268591 & 1.77912994761774e+01 \\
268591 & 1.77912994046798e+01 \\
\end{tabular}
\end{center}
\caption{Precision of eigenvalues with changing mesh refinement and repeated eigenvalue solutions with 3 different guesses for the initial eigenvector.}
\label{precision.table}
\end{table}

\begin{figure}
    \begin{center}
  \includegraphics[width=\columnwidth]{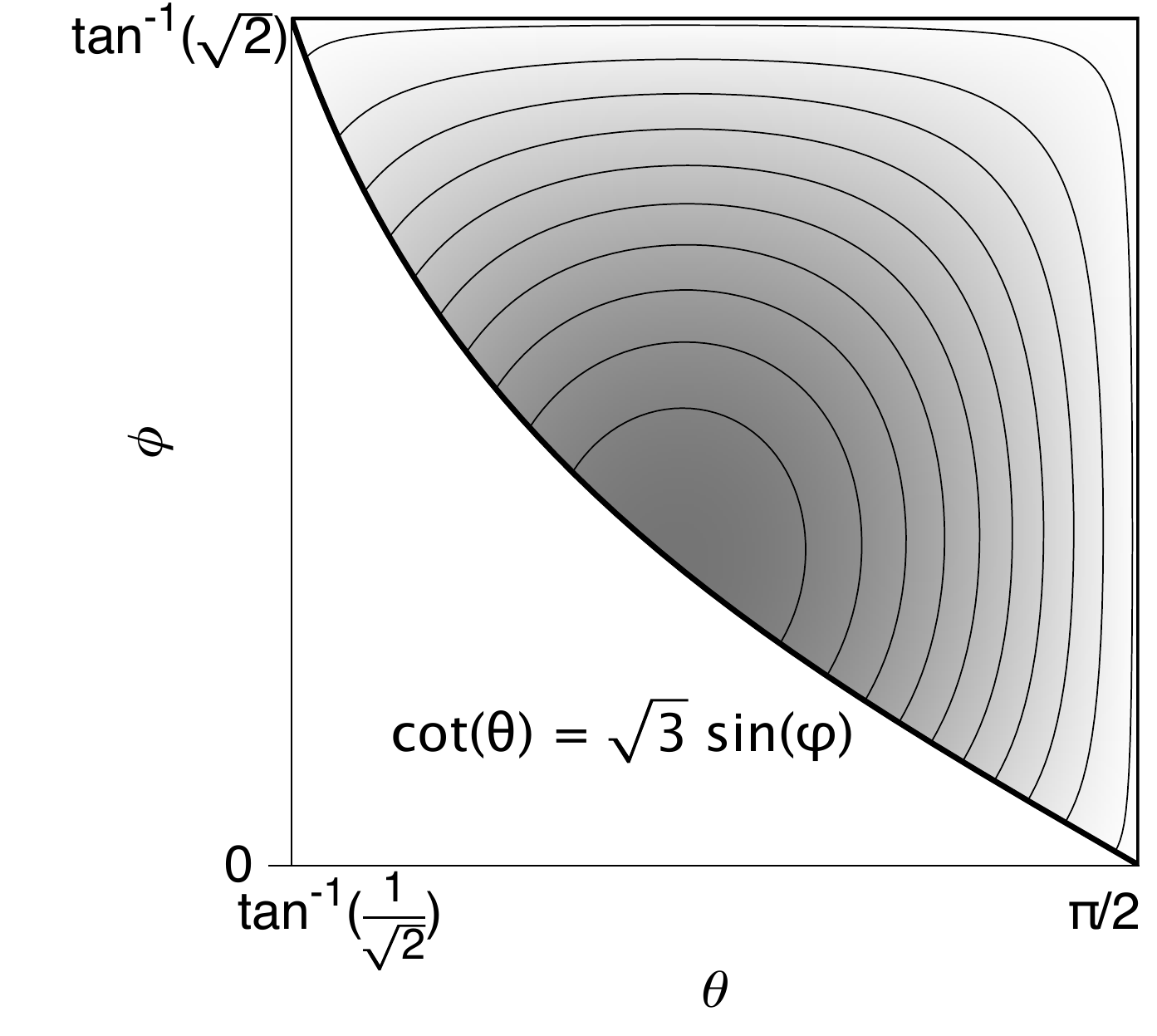}%
\hfill
  \includegraphics[width=\columnwidth]{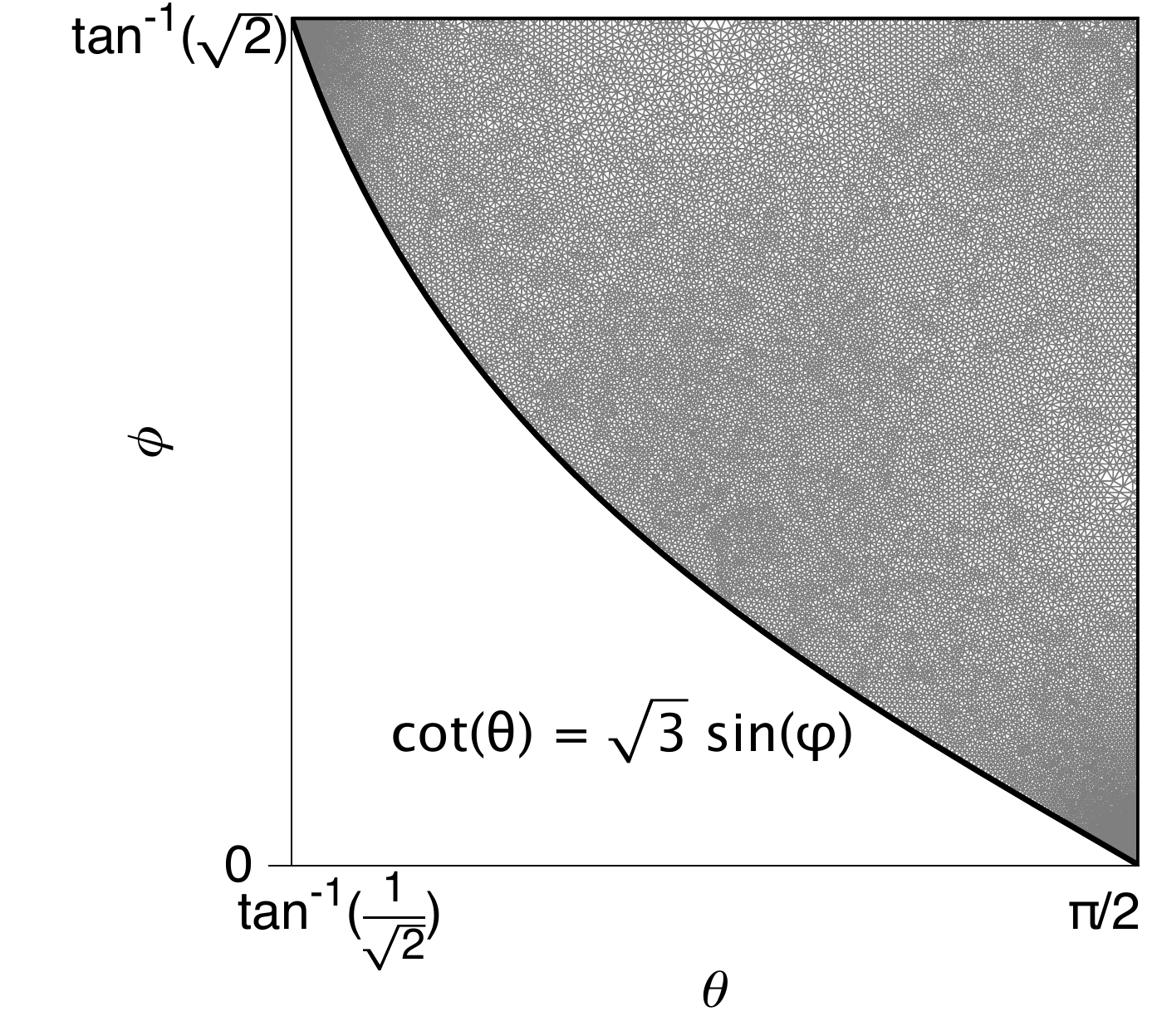}%
\end{center}
%
%
    \caption{Eigenfunctions obtained with the numerical solver for case~C of Table~\ref{exponents.table} (top) and intermediate resolution adapted finite element mesh~(bottom).  The mesh is visible only upon enlargement (of the electronic version).
We use the basic triangular domain of one particular ordering (e.g., `1234'), 
and symmetry is exploited in the analysis of each case by imposing appropriate boundary conditions: for our case~C, Dirichlet  conditions on the top and right sides, and  a Neumann condition on the curved side.
}  
\label{eigen.fig}
\end{figure}

Figure~\ref{eigen.fig} shows the final estimate of the eigenfunction solution for this case and mesh.  Although difficult to see without enlarging the figure, the mesh is not uniform in resolution.  The resolution increases in the top left and bottom right corners of the domain because the second derivative goes to infinity there.  This singular behavior makes obtaining accurate solutions difficult using a uniform mesh resolution.   The cause of the singularity is an incompatibility between the boundary conditions;  along the straight sides of the domain, a homogeneous Dirichlet boundary condition is enforced which means that isocontours of the eigenfunction must be parallel to these surfaces.  On the curved boundary, a homogeneous Neumann boundary condition is enforced which means that isocontours must be perpendicular to the curved side.   Because the curved surface is not perpendicular to the straight surfaces at their intersection points, it is impossible to satisfy both conditions.  This drives an infinite second derivative in the eigenfunction as the two intersection points are approached.   The mesh adaption routine detects the increased truncation error that occurs at these points and refines the mesh near the corners.  For the final mesh for this case, the ratio of the largest to smallest edge length in the mesh is 100.  


%

\section{Results}

Enumerating all the possible ordering statistics for $N=4$ is already in itself a non-trivial problem (that we do not fully attempt here).  Instead, we focus on ordering statistics that have been studied to date (e.g., leader, laggard) as well as some ``new" statistics that seem to illustrate a specific point.   Our findings are summarized in Table~\ref{exponents.table}.  

The first and simplest case (case A), where the single walker is confined to the `1234' wedge, corresponds to the problem of vicious walkers.  The solution, $\mu=7$ and $\b=(\mu-1)/2=3$, is known exactly (through the method of images).
We ran the numerical solver for this case as a test,  finding agreement with the exact result to better than 8 digits.  As described earlier, the $S$-domain corresponding to a single wedge is the isosceles right angle triangle, on the unit sphere, shown in Fig.~\ref{eigen.fig} (with Dirichlet boundary conditions on all three sides).
  It is easiest to describe all the other cases in terms of combinations of this basic triangle.  The results in Table~\ref{exponents.table} are arranged in order of the number of basic triangles involved for each statistics, or equivalently, by the fraction $\Omega/4\pi$ of the total solid angle covered by the domain $S$ of each case (last column).  A straightforward prediction of the cone approximation~\cite{ben-naim10a,ben-naim10b} is that $\mu$ and $\beta$ get smaller the larger the fraction of the solid angle covered.

There are two distinct combinations of 2 basic triangles, depending on wether they join along one of the legs or along the hypotenuse.  The former, case~B, corresponds to the adjacent orderings `1234' and `1243' (and symmetric variations). We write this ordering statistics symbolically as `12$\bullet$$\bullet$', the bullets denoting the fact that particles 3 and 4 are free to exchange positions. The other option, case~C, of joining two triangles through an hypotenuse --- `1234' and `1324', for example --- yields the ``bookends" statistics  `1$\bullet$$\bullet$4': particles 1 and 4 retain their first and last position, respectively, bookending particles 2 and 3 in between (which may cross one another).  While the domains for both cases B and C span the same solid angle ($1/12$ of the total sphere), the results for $\mu$ and $\beta$ are somewhat different, with the higher values corresponding to the more irregular, or more elongated shape of the domain for case~B. (The smallest possible values of $\beta$ and $\mu$ would be achieved for a circular domain $S$, the one assumed in the cone approximation~\cite{ben-naim10a,ben-naim10b}.)

\begin{table}
\centering
\begin{tabular}{|c|c|ccc|}
\hline
case & ordering statistics & $\beta$ & $\mu$ & $\Omega/4\pi$ \\
\hline\hline
A & vicious walkers (1234) &3 (exact) & 7 (exact) & 1/24 \\
B & 12$\bullet$$\bullet$ & 2.0716054 & 5.1432108 & 1/12 \\
C & ``bookends" (1$\bullet$$\bullet$4) & 1.8737525 & 4.7475050 & 1/12 \\
D & 1 leads, and 2 ahead of 3 & 1.6204515 & 4.2409030 & 1/8 \\
E & ``teams" ($\bullet$$\bullet$$|$$\bullet$$\bullet$) & 1.1949400 & 3.3898800 & 1/6 \\
F & leader (1$\bullet$$\bullet$$\bullet$) & 0.91287850 & 2.8257570 & 1/4 \\
G & 1\&2 or 1\&3 team-lead & 0.93265225 & 2.8653045 & 1/3 \\
H & 1 leads or 2 leads & 0.61257504 & 2.2251500 & 1/2 \\
J & 1 in 1st or 2nd place & 0.55480541 & 2.1096108 & 1/2 \\
K & team excluded from edges & 0.81433951 & 2.6286790 & 2/3 \\
L & laggard (2 never leads) & 0.30763604 & 1.6152721 & 3/4\\
\hline
\end{tabular}
\caption{Numerical results for the various cases discussed in the text.  The probability to maintain the particular ordering in each case decays as $t^{-\b}$.  Alternatively, the electrostatic potential within the $W$-domain corresponding to that case falls off as $r^{-\mu}$ at large $r$, and the  relation $\b=(\mu-1)/2$ holds (for $N=4$).  The entries are arranged by decreasing order of the solid angle $\Omega$ sustained by the domain,~$W$ (last column).  The well known exact result for vicious walkers (case~A)~\cite{fisher84,huse84} has been included for completeness.}
\label{exponents.table}
\end{table}

There are two ways to join 3 adjacent triangles but we consider only the more regular-shaped case~D, consisting for example of the orderings `1234', `1324' and `1342' (the neglected possibility has a concave-shaped domain).  There is no simple way to describe this ordering statistics, besides perhaps that particles 1, 3, and 4 retain their relative ordering, while particle 2 can be anywhere but never ahead of 1.  A particular interest in this case arises from the fact that its domain is {\em exactly} one-half of the domain for the leader problem (case~F).  Thus, the first (lowest) mode for case~D corresponds to the second mode of case~F, yielding the leading correction for the latter (see below).

There are six ways to join 4 basic triangles and again we consider only the most compact-shaped possibility, case~E, consisting of the orderings making one face of the cube in Fig.~\ref{n4.fig}, for example, `1234', `1243',
`2143', and `2134'. In this case particles 1 and 2 ``team-lead" --- they occupy the first and second position but not necessarily in that order.  Particles 3 and 4 ``team-lag", occupying the last two positions (without  regard to their relative ordering).  We denote this, symbolically, as `$\bullet$$\bullet$$|$$\bullet$$\bullet$'.

Of the many statistics available for 6 adjacent orderings we focus on the popular leader problem, case~F.  The domain $S$ for particle $i$ to lead consists of all of the six orderings surrounding the vertex labeled `$i$' in Fig~\ref{n4.fig}b (see caption). The best approximation available to date for the corresponding exponent is $\b=0.913$~\cite{dba03,ben-naim10a}.  In fact, more than three digits accuracy are claimed in~\cite{dba03} but the extra digits are inconsistent with our present findings.  The result in~\cite{dba03} relies on an extrapolation that assumes $r^{-4}$ for the leading correction of $V(\r)\sim r^{-\mu}$, instead
of the more accurate leading correction  $\sim r^{-4.2409030}$ suggested by our case~D, and this may partly explain the discrepancy. 

In case~G either particles 1 and 2 team-lead or particles 1 and 3 team-lead, at any given time.  The domain consists of all 8 orderings in the two faces visible at the bottom of the cube in Fig~\ref{n4.fig}b.
An interesting fact is that the $\b$ and $\mu$ exponents are not smaller than those for the leader problem, even though the fraction of solid angle covered is larger ($1/3$, as opposed to $1/4$ for the leader).  This may be the result of the more irregularly-shaped (elongated) domain of case~G.

Case~H describes the probability that at any given time either particle 1 leads or particle 2 leads (alternatively,
particles 3 and 4 never lead, or lag jointly).  The domain $S$ consists of the union of the domains for `particle 1 leads' and `particle 2 leads', comprising of 12 basic triangles and spanning a solid angle of $2\pi$, or $1/2$ of the sphere.  Case~J concerns the ordering statistics for particle $i$ never to fall below the second place.  The domain encompasses all of the 12 orderings visible in Fig~\ref{n4.fig}b (for the choice of $i=1$).  The exponents for both cases are somewhat larger than $\mu=2$ and $\b=1/2$ predicted by the cone approximation~\cite{ben-naim10a,ben-naim10b}.  This, and the fact that the exponents for case~H are larger than those for case~J is explained from the irregularity of the domains (domain J more irregular than H, which in turn is more irregular than the half sphere). 
Ben-Naim and Krapivsky~\cite{ben-naim10a} studied the probability that in an $N$-walkers field a particle doesn't fall below the $n$-th place.  Our case~J is quite in agreement with their simulations result of $\b=0.556$, for $N=4$ and $n=2$.

Case~K describes the statistics for a team of particles to be excluded from the edges.  For example, particles 1 and 2 are not allowed to team-lead (occupy the first two positions) nor to team-lag (occupy the last two positions).  The domain consists of 4 faces on the envelope of the cube in Fig.~\ref{n4.fig}b: the bottom-left face where 1 and 2 team-lead, and the opposite (non-visible) face where 1 and 2 team-lag are excluded.  The exponents for this ring-shaped, highly irregular domain are quite larger than expected from the solid angle covered ($2/3$ of the sphere) and almost on par with the exponents for case~G whose domain is only half as large ($1/3$ of the sphere).

Finally, case~L is that of the laggard.  The domain comprises of  all of the orderings where particle $i$ does {\em not} lead, covering $3/4$ of the solid angle of the sphere.  Our results here are consistent with $\b=0.30$~\cite{dba03} and $\b=0.306$~\cite{ben-naim10a} obtained from numerical simulations in previous studies.  The fact
that $\b_{\rm laggard}\approx(1/3)\b_{\rm leader}$ (cases~L and F, respectively) is intriguing and we look into this next.

\subsection*{Correlations and $n$ out of $N$ laggards}
In the general case of $N$ walkers, the probability that particle 1 leads to time $t$ equals the joint probability
that particles $2, 3,\dots,N$ lag, that is, that none of them  leads to time $t$;
\begin{equation}
\Plead{1}=\Plag{2,3,\dots,N}\;.
\end{equation}
If correlations could be ignored, the joint probability for the laggards would simplify to
\begin{equation}
\begin{split}
\Plag{2,3,\dots,N}&=\Plag{2}\Plag{3}\cdots\Plag{N}\\
&=\Plag{2}^{N-1},
\end{split}
\end{equation}
the last equation resulting from the fact that all particles are equally likely to lag.
Denoting the ordering exponents for leading and lagging as $\Plead{1}\sim t^{-\blead}$ and $\Plag{2}\sim t^{-\blag}$, we then get
\begin{equation}
\blead=(N-1)\blag\,.
\end{equation}
This seems to be the case for $N=3$, where $\blead=3/4$, $\blag=3/8$, and $\blead=2\blag$, but it cannot be generally true; in the limit of $N\to\infty$, for example, $\blead\sim(\ln N)/4 < (N-1)\blag\sim\ln N$.  Are there no correlations in the case of $N=3$, and if so, at what $N$ do correlations creep in? --- Our results show quite convincingly that correlations creep in already for $N=4$.  Indeed, using the exponents for cases F (leader) and 
L (laggard) of Table~\ref{exponents.table}, we get $\blead=0.91287850<3\blag=.92290812$, although the difference is tiny (it could barely be intimated from the previously~\cite{ben-naim10a} best available estimate of $\blag=0.306$).

We can, in fact, probe correlations somewhat more deeply.  The probability that particle 1 leads can be written {\em exactly} as
\begin{equation}
\begin{split}
\Plead{1}=&\Plag{2}\Plag{3|2}\Plag{4|2,3}\\
&\cdots \Plag{N|2,\dots,N-1}\,,
\end{split}
\end{equation}
where we have used the notation
\begin{equation}
\Plag{i|j,k,\dots,n}=\frac{\Plag{i,j,k\dots,n}}{\Plag{j,k,\dots,n}}
\end{equation}
for the conditional probability that particle $i$ lags, given that particles $j,k,\dots,n$ lag as well.  In the absence of correlations all these  conditional probabilities are equal to $\Plag{2}$, as is the case for $N=3$.  For $N=4$, we
obtain the pertinent ordering exponents from Table~\ref{exponents.table}: $\Plag{2}$ is case~L, $\Plag{2,3}$ is case~H, and $\Plag{2,3,4}$ is the same as $\Plead{1}$, or case~F.  We thus obtain,
\[
\Plag{3|2}\sim t^{-0.30493900}{\rm\ and\ } \Plag{4|2,3}\sim t^{-0.30030346}.
\]
The differences from $\Plag{2}\sim t^{-0.30763604}$ are tiny and show only with better than three digits accuracy, but the effect is undeniable.

That the correlations are small can be understood intuitively from the following argument.  If we have the constraint that particles $j,k,\dots,n$ lag, it might mean one of two things: either these particles are unusually ``slow" (hence they lag), or some of the remaining particles are unusually ``fast" (leaving them behind).  In the first case, it'd be more difficult for particle $i$ to compete with the laggards and lag as well, whereas in the second case it'd be easier for it to lag, in comparison to the fast particles...  These two contradicting tendencies seem to balance out for $N=3$, but our numerical results suggest that for $N=4$ it is somewhat easier to lag given the constraint that one other particle is lagging, and  easier still if two other particles are lagging.  It is plausible that this trend is general for $N\geq4$, that is, that $\Plag{i|j,k,\dots,n}$ is larger the greater $n$ is, but we are unable to prove this notion.
Finally, we note that the delicate balance for the case of $N=3$ seems to be accidental; indeed correlations clearly arise as soon as the particles are given slightly different diffusion constants~\cite{dba03}.

\section{Discussion}

In summary, we have provided numerical estimates, accurate to 8 digits, for the decay exponent $\beta$ of various ordering statistics for $N=4$ random walkers on the line.  The results were found by examining an analogous problem in electrostatics; the actual numerical computations yield an estimate for the exponent $\mu$ in $V(\r)\sim r^{-\mu}$, for the long-range decay ($r\to\infty$) of the electric potential $V$ within three-dimensional wedges with absorbing boundary conditions ($V=0$ on the walls of the wedges).  Employing the ansatz $V(r,\theta,\phi)\sim r^{-\mu}f(\theta,\phi)$ results in an eigenvalue equation for $f(\theta,\phi)$ and the eigenvalue $\mu$ can be then found numerically with great accuracy, since the domain for $f$ is finite (in contrast to the infinite domain for the problem in $V$).  The same technique has been used before for $N=3$ walkers, as reviewed in Section~\ref{methods.sec}.

Some of the domains that we have considered display a high degree of symmetry.  For example, the domain for the leader problem is a tetrahedral wedge with a solid angle of exactly $1/4$ of the sphere (the laggard problem
involves the complementary space, outside of this domain); the ``teams" ordering statistics (case~E) involves a square pyramidal wedge, with a solid angle of $1/6$ of the sphere, etc.  To our surprise and despite our best efforts, we have failed to find analytical solutions to these problems in the literature.  An interesting case in point (but unrelated to $N=4$ walkers) is that of the cartesian corner $x,y,z>0$.  If the walls are absorbing (or $V=0$), the potential inside such a domain falls off as $V(\r)\sim r^{-4}$, and a random walker that dies on the walls survives
to time $t$ with probability $P(t)\sim t^{-3/2}$, as can be found exactly through the method of images.  But what about the complementary domain? --- What if the walker or the electric charge reside {\em outside} of the first octant?  Using our numerical techniques we were able to estimate that $V\sim r^{-1.45417}$ and $P\sim t^{-0.227086}$ (to 6 significant digits) but we were unable to track an analytic answer in the literature even for this seemingly simple case.  

For $N=3$ the probability that particles 2 and 3 lag jointly (neither of them ever becomes the leader) are uncorrelated in the long-time limit, equaling the square of the probability that a single particle lags.  In contrast, our numerical analysis shows that for $N=4$ the simultaneous lagging of two or three particles {\em is} correlated, although the correlations are small and manifest only in the third digit of the corresponding probability-decay exponents.  The more general problem of the probability that $n$ out of $N$ particles lag simultaneously (none of the $n$ particles leads, to time $t$) might be of interest.  In particular, it might be nice to establish how the decay exponents scale with $n$ and $N$, in a similar fashion to that obtained by Ben-Naim and Krapivsky~\cite{ben-naim10a,ben-naim10b} for other ordering statistics.

Finally, we have focused on only the small subset of ordering statistics summarized in Table~\ref{exponents.table}, without even attempting to count how many cases we have left out.  Enumerating the distinct types of ordering statistics for $N\geq4$ (in the restricted sense of the present work) remains an interesting combinatorial problem.


\bibliographystyle{apsrev4-1}

\bibliography{_mainbib.bib,slepc.bib,dani.bib,_hp.bib}

\end{document}